\documentclass[11pt,fleqn]{article}
\usepackage{amsfonts, epsfig, amssymb}
\usepackage{color}
\newcommand{\ol}{\setlength{\itemsep}{0pt.}\begin{enumerate}}
\newcommand{\eol}{\end{enumerate}\setlength{\itemsep}{-\parsep}}
\newcommand{\ignore}[1]{}
\title{The "Most informative boolean function" conjecture holds for high noise}
\author{Alex Samorodnitsky}

\begin{document}
\date{}
\maketitle


\newtheorem{THEOREM}{Theorem}[section]
\newenvironment{theorem}{\begin{THEOREM} \hspace{-.85em} {\bf :}
}%
                        {\end{THEOREM}}
\newtheorem{LEMMA}[THEOREM]{Lemma}
\newenvironment{lemma}{\begin{LEMMA} \hspace{-.85em} {\bf :} }%
                      {\end{LEMMA}}
\newtheorem{COROLLARY}[THEOREM]{Corollary}
\newenvironment{corollary}{\begin{COROLLARY} \hspace{-.85em} {\bf
:} }%
                          {\end{COROLLARY}}
\newtheorem{PROPOSITION}[THEOREM]{Proposition}
\newenvironment{proposition}{\begin{PROPOSITION} \hspace{-.85em}
{\bf :} }%
                            {\end{PROPOSITION}}
\newtheorem{DEFINITION}[THEOREM]{Definition}
\newenvironment{definition}{\begin{DEFINITION} \hspace{-.85em} {\bf
:} \rm}%
                            {\end{DEFINITION}}
\newtheorem{EXAMPLE}[THEOREM]{Example}
\newenvironment{example}{\begin{EXAMPLE} \hspace{-.85em} {\bf :}
\rm}%
                            {\end{EXAMPLE}}
\newtheorem{CONJECTURE}[THEOREM]{Conjecture}
\newenvironment{conjecture}{\begin{CONJECTURE} \hspace{-.85em}
{\bf :} \rm}%
                            {\end{CONJECTURE}}
\newtheorem{MAINCONJECTURE}[THEOREM]{Main Conjecture}
\newenvironment{mainconjecture}{\begin{MAINCONJECTURE} \hspace{-.85em}
{\bf :} \rm}%
                            {\end{MAINCONJECTURE}}
\newtheorem{PROBLEM}[THEOREM]{Problem}
\newenvironment{problem}{\begin{PROBLEM} \hspace{-.85em} {\bf :}
\rm}%
                            {\end{PROBLEM}}
\newtheorem{QUESTION}[THEOREM]{Question}
\newenvironment{question}{\begin{QUESTION} \hspace{-.85em} {\bf :}
\rm}%
                            {\end{QUESTION}}
\newtheorem{REMARK}[THEOREM]{Remark}
\newenvironment{remark}{\begin{REMARK} \hspace{-.85em} {\bf :}
\rm}%
                            {\end{REMARK}}

\newcommand{\thm}{\begin{theorem}}
\newcommand{\lem}{\begin{lemma}}
\newcommand{\pro}{\begin{proposition}}
\newcommand{\dfn}{\begin{definition}}
\newcommand{\rem}{\begin{remark}}
\newcommand{\xam}{\begin{example}}
\newcommand{\cnj}{\begin{conjecture}}
\newcommand{\mcnj}{\begin{mainconjecture}}
\newcommand{\prb}{\begin{problem}}
\newcommand{\que}{\begin{question}}
\newcommand{\cor}{\begin{corollary}}
\newcommand{\prf}{\noindent{\bf Proof:} }
\newcommand{\ethm}{\end{theorem}}
\newcommand{\elem}{\end{lemma}}
\newcommand{\epro}{\end{proposition}}
\newcommand{\edfn}{\bbox\end{definition}}
\newcommand{\erem}{\bbox\end{remark}}
\newcommand{\exam}{\bbox\end{example}}
\newcommand{\ecnj}{\bbox\end{conjecture}}
\newcommand{\emcnj}{\bbox\end{mainconjecture}}
\newcommand{\eprb}{\bbox\end{problem}}
\newcommand{\eque}{\bbox\end{question}}
\newcommand{\ecor}{\end{corollary}}
\newcommand{\eprf}{\bbox}
\newcommand{\beqn}{\begin{equation}}
\newcommand{\eeqn}{\end{equation}}
\newcommand{\wbox}{\mbox{$\sqcap$\llap{$\sqcup$}}}
\newcommand{\bbox}{\vrule height7pt width4pt depth1pt}
\newcommand{\qed}{\bbox}
\def\sup{^}

\def\H{\{0,1\}^n}

\def\S{S(n,w)}

\def\g{g_{\ast}}
\def\xop{x_{\ast}}
\def\y{y_{\ast}}
\def\z{z_{\ast}}

\def\f{\tilde f}

\def\n{\lfloor \frac n2 \rfloor}

\def \E{\mathop{{}\mathbb E}}
\def \R{\mathbb R}
\def \Z{\mathbb Z}
\def \F{\mathbb F}
\def \T{\mathbb T}

\def \x{\textcolor{red}{x}}
\def \r{\textcolor{red}{r}}
\def \Rc{\textcolor{red}{R}}

\def \noi{{\noindent}}

\def \iff{~~~~\Leftrightarrow~~~~}

\def\<{\left<}
\def\>{\right>}
\def \({\left(}
\def \){\right)}

\def \e{\epsilon}
\def \l{\lambda}

\def\Tp{Tchebyshef polynomial}
\def\Tps{TchebysDeto be the maximafine $A(n,d)$ l size of a code with distance $d$hef polynomials}
\newcommand{\rarrow}{\rightarrow}

\newcommand{\larrow}{\leftarrow}

\overfullrule=0pt
\def\setof#1{\lbrace #1 \rbrace}

\begin{abstract}
\noi We prove the "Most informative boolean function" conjecture of Courtade and Kumar for high noise $\e \ge 1/2 - \delta$, for some absolute constant $\delta > 0$.

\noi Namely, if $X$ is uniformly distributed in $\H$ and $Y$ is obtained by flipping each coordinate of $X$ independently with probability $\e$, then, provided $\e \ge 1/2 - \delta$, for any boolean function $f$ holds $I\Big(f(X);Y\Big) \le 1 - H(\e)$. This conjecture was previously known to hold only for balanced functions \cite{NELAPP}.
\end{abstract}

\section{Introduction}

\noi We start with recalling the conjecture of Courtade and Kumar \cite{Kumar-Courtade}.

\noi Let $(X,Y)$ be jointly distributed in $\H$ such that their marginals are uniform and $Y$ is obtained by flipping each coordinate of $X$ independently with probability $\e$. Let $H$ denote the binary entropy function $H(x) = -x \log_2 x - (1-x) \log_2(1-x)$. The conjecture of \cite{Kumar-Courtade} is:

\cnj
\label{cnj:1}
For all boolean functions $f:~\H \rarrow \{0,1\}$,
\[
I\Big(f(X);Y\Big) \quad \le \quad 1 ~-~ H(\e)
\]
\ecnj

\noi This inequality holds with equality if $f$ is a characteristic function of a subcube of dimension $n-1$. Hence, the conjecture is that such functions are the "most informative" boolean functions.

\noi This note is a follow-up to the paper \cite{NELAPP}, in which the conjecture was shown to hold for $\e$ close to $1/2$ and for balanced boolean functions $f$. Here we make a few simple modifications to the argument, in order to remove the requirement on the boolean functions to be balanced. This note is not self-contained and we suggest that it should be read as an addendum to \cite{NELAPP}. In particular, we use the notation from that paper.

\noi Our main result is
\thm
\label{thm:main}
There exists an absolute constant $\delta > 0$ such that for any noise $\e \ge 0$ with $(1-2\e)^2 \le \delta$ and for any boolean function $f:~\H \rarrow \{0,1\}$
holds
\[
I\Big(f(X);Y\Big) \quad \le \quad 1 ~-~ H(\e)
\]
\ethm

\noi The argument proceeds similarly to \cite{NELAPP}. We first argue, following \cite{OSW}, that if a boolean function $f$ is 'informative', meaning that $I\Big(f(X);Y\Big) ~ \ge ~ 1 - H(\e)$, then its Fourier mass is concentrated on the first two levels. By the inequality of \cite{FKN} (see \cite{FKN}, Theorem 1.1) this implies that $f$ is close to a characteristic function of a subcube. As the last step, we use a modified version of a theorem from \cite{NELAPP} to show that for such functions the conjecture holds.

\noi {\bf Notation} (See \cite{NELAPP}): For a function $f$ on $\H$ we write $f = \sum_{S \subseteq [n]} \widehat{f}(S) \cdot W_S$ for the Fourier expansion of $f$. For a nonnegative function $f$, we let $Ent\Big(f\Big)$ be the entropy of $f$. For $0 \le \e \le 1/2$, we denote by $T_{\e}$ the appropriate noise operator. We note that (\cite{NELAPP}) for a boolean function $f:~\H \rarrow \{0,1\}$ holds

\beqn
\label{I-ent}
I\Big(f(X);Y\Big) \quad = \quad Ent\Big(T_{\e} f\Big) ~+~ Ent\Big(T_{\e} (1-f)\Big)
\eeqn

\noi We write $\l ~=~ (1-2\e)^2$ and recall that
\beqn
\label{series}
1 - H(\e) ~~=~~ \frac{1}{\ln 2} \cdot \sum_{k=1}^{\infty} \frac{1}{2k(2k-1)} \cdot \l^k
\eeqn

\noi {\bf Brief overview}. The main issue here is dealing with non-balanced functions. We prove two technical claims.

\noi First, we slightly extend the approach of \cite{OSW} and show that for any nonnegative non-zero function $f$ holds\footnote{In this paper asymptotic notation hides absolute constants independent of the remaining parameters.}
\lem
\label{lem:first level}
\[
Ent\Big(T_{\e} f\Big) ~~ \le ~~ \frac{1}{\E f} \cdot \(\frac{1}{2\ln 2} \cdot \sum_{k=1}^n \widehat{f}^2(\{k\})\) \cdot \l  ~+~  O\(\frac{\E f^2}{\E f} \cdot \l^{4/3}\) ~+~  O\(\frac{\E^2 f^2}{\E^3 f} \cdot \l^2\)
\]
\elem

\noi Second, we show that the proof of Theorem~1.11 in \cite{NELAPP} can be modified to give:

\thm
\label{thm:KC-ln}
There exists an absolute constant $\delta > 0$ such that for any noise $\e \ge 0$ with $(1-2\e)^2 \le \delta$ and for any boolean function $f:~\H \rarrow \{0,1\}$ such that
\begin{itemize}
\item

$\frac12 - \delta \le \E f \le \frac12$;

\item

There exists $1 \le k \le n$ such that $|\widehat{f}(\{k\})| \ge (1 - \delta) \cdot \E f$

\end{itemize}

\noi Holds
\[
Ent\Big(T_{\e} f\Big) ~~+~~ Ent\Big(T_{\e} (1-f) \Big)\quad \le \quad 1 - H(\e)
\]
\ethm

\noi Theorem~\ref{thm:main} is an easy corollary of these two claims and the result of \cite{FKN}.

\noi {\bf Organization}. This paper is organized as follows. We deduce Theorem~\ref{thm:main} from Lemma~\ref{lem:first level} and Theorem~\ref{thm:KC-ln} in Section~\ref{sec:main}. We prove Lemma~\ref{lem:first level} in Section~\ref{sec:lem}, and Theorem~\ref{thm:KC-ln} in Section~\ref{sec:thm}.

\section{Proof of Theorem~\ref{thm:main}}
\label{sec:main}

\noi It is known (see \cite{Kumar-Courtade}) that for any boolean function $f$ holds $I\(f(X);Y\) ~\le~ \l \cdot H\(\E f\)$, where $H$ is the binary entropy function. This immediately implies the validity of Conjecture~\ref{cnj:1} for boolean functions with expectation lying in $[0,c] ~\cup~ [1-c,1]$, for some absolute constant $0 < c < 1/2$.

\noi In addition, we may assume, by symmetry, that $\E f \le 1/2$. Combining these two observations, it remains to consider the case
\beqn
\label{exp-bal}
c ~~\le~~ \E f ~~\le~~ 1/2
\eeqn

\noi Let $f$ be a boolean function with $I\(f(X);Y\) ~\ge~  1 - H(\e)$. By (\ref{I-ent}) this is the same as
\[
Ent\Big(T_{\e} f\Big) ~~+~~ Ent\Big(T_{\e} (1-f)\Big) \quad \ge \quad 1 ~-~ H(\e)
\]

\noi On the other hand, applying Lemma~\ref{lem:first level} to the functions $f$ and $1-f$ and taking into account (\ref{exp-bal}) gives
\[
Ent\Big(T_{\e} f\Big) ~+~ Ent\Big(T_{\e} (1-f)\Big) ~~\le~~ \frac{1}{\E f \Big( 1- \E f\Big)} \cdot \(\frac{1}{2\ln 2} \cdot \sum_{k=1}^n \widehat{f}^2(\{k\})\) \cdot \l  ~+~ O\Big(\l^{4/3}\Big)
\]

\noi Combining these two inequalities and observing that (\ref{series}) implies $1 - H(\e) ~\ge~ \frac{\l}{2 \ln 2}$ shows
\[
\sum_{k=1}^n \widehat{f}^2(\{k\}) \quad \ge \quad \E f \cdot \Big(1 - \E f\Big) ~-~ O\(\l^{1/3}\)
\]
Recall that for a boolean function $f:\H \rarrow \{0,1\}$ holds $\E f^2 = \E f$, and hence $\sum_{S \not = \emptyset} \widehat{f}^2(S) ~=~ \E f ( 1 - \E f)$. This means that the preceding inequality implies $\sum_{|S| \ge 2} \widehat{f}^2(S) ~\le~ O\(\l^{1/3}\)$.

\noi We now proceed similarly to the proof of Theorem~1.11 in \cite{NELAPP}. If $\l$ is sufficiently small, the application of the inequality of \cite{FKN}, or of its a precise version due to \cite{JOW} (see Theorem~5.5 in \cite{NELAPP}), and taking into account (\ref{exp-bal}), imply that $f$ meets the conditions of Theorem~\ref{thm:KC-ln}, and hence Conjecture~\ref{cnj:1} holds for $f$.

\section{Proof of Lemma~\ref{lem:first level}}
\label{sec:lem}

\noi We may and will assume, by homogeneity, that $\E f = 1$.

\noi Let us introduce some notation. For $x \in \H$, let $x^c$ be the complement of $x$, that is the element of $\H$ with $x^c_i = 1 - x_i$ for all $1 \le i \le n$.

\noi For a nonnegative function $g$ on $\H$, let $g_0$ be the 'even' part of $g$ defined by $g_0(x) ~=~ \(g(x) + g\(x^c\)\)/2$, and let $g_1 ~=~ g - g_0$ be the 'odd' part of $g$. By definition, $g_0(x) ~=~ g_0\(x^c\)$ and $g_1(x) ~=~ -g_0\(x^c\)$. Note also that $|g_1| \le g_0$.

\noi We will need the following well-known (and easy to verify) fact:
\[
g_0 ~~=~~ \sum_{|S| ~even} \widehat{g}(S) \cdot W_S \quad \mbox{and} \quad g_1 ~~=~~ \sum_{|S| ~odd} \widehat{g}(S) \cdot W_S
\]

\noi We start with an auxiliary claim.
\lem
\label{lem:aux}
For any nonnegative function $g$ with expectation $1$ holds
\[
Ent\(g\) ~~=~~ Ent\Big(g_0\Big) ~+~ \E_x ~g_0(x) \cdot \Bigg(1 - H\(\frac{1 - |g_1(x)|/g_0(x)}{2}\)\Bigg)
\]
Here for $x$ such that $g_0(x) = g_1(x) = 0$, the expression $g_0(x) \cdot \(1 - H\(\frac{1 - |g_1(x)|/g_0(x)}{2}\)\)$ is interpreted as $0$.
\elem

\prf
We have
\[
Ent\Big(g\Big) \quad = \quad \E_x g(x) \log g(x) \quad = \frac12 \cdot \E_x \Big( g(x) \log g(x) ~+~ g\(x^c\) \log \(x^c\) \Big) \quad =
\]
\[
\frac12 \cdot \E_x \bigg(\Big(g_0(x) + g_1(x) \Big) \cdot \log \Big(g_0(x) + g_1(x) \Big) ~~+~~ \Big(g_0(x) - g_1(x) \Big) \cdot \log \Big(g_0(x) - g_1(x) \Big)\bigg)
\]
It is easy to verify that for any $0 \le b \le a$ holds \\
$1/2 \cdot \Big((a+b) \log(a+b) + (a-b) \log(a-b)\Big) ~=~ a \log a + a \cdot \(1 - H\(\frac{1 - b/a}{2}\)\)$, where the last expression should be interpreted as $0$ for $a = b = 0$.

\noi Using this identity with $a = g_0(x)$ and $b = g_1(x)$ gives the claim of the lemma.
\eprf

\noi Next, as in \cite{OSW}, we upper bound $1 - H\(\frac{1-x}{2}\)$ by $\frac{1}{2 \ln 2} \cdot x^2 + \(1 - \frac{1}{2 \ln 2}\) \cdot x^4$.

\noi This gives
\[
Ent\(g\) ~~\le~~ Ent\Big(g_0\Big) ~~+~~ \frac{1}{2 \ln 2} \cdot \E_x ~\frac{g^2_1(x)}{g_0(x)} ~~+~~ \(1 - \frac{1}{2 \ln 2}\) \cdot \E_x ~\frac{g^4_1(x)}{g^3_0(x)}
\]

\noi Substitute $g ~=~ T_{\e} f$. It is easy to verify $\Big(T_{\e} f\Big)_i ~=~ T_{\e} \Big(f_i\Big)$ for any function $f$ and for $i = 0, 1$. Consequently:
\[
Ent\Big(T_{\e} f\Big) ~~\le~~ Ent\Big(T_{\e} f_0\Big) ~+~ \frac{1}{2 \ln 2} \cdot \E_x ~\frac{\Big(T_{\e} f_1(x)\Big)^2}{T_{\e} f_0(x)} ~+~ \(1 - \frac{1}{2 \ln 2}\) \cdot \E_x ~\frac{\Big(T_{\e} f_1(x)\Big)^4}{\Big(T_{\e} f_0(x)\Big)^3}
\]

\noi We upper bound each of the summands on the RHS separately.

\begin{enumerate}

\item
The first summand. Note that $\E T_{\e} f_0 ~=~ \E f_0 ~=~ \E f ~=~ 1$. Hence, by Lemma~5.4 in \cite{NELAPP},
\[
Ent\Big(T_{\e} f_0\Big) \quad \le \quad O\bigg(\sum_S |S| \cdot \widehat{T_{\e} f_0}^2(S) \bigg) \quad = \quad O\bigg(\sum_S |S| \l^{|S|} \widehat{f_0}^2(S) \bigg) \quad =
\]
\[
O\bigg(\sum_{|S|~even} ~|S| \l^{|S|} \widehat{f}^2(S) \bigg) \quad = \quad O\Big(\E f^2 \cdot \l^2\Big)
\]

\item
The second summand. First we argue that $T_{\e} f_0$ is bounded away from $0$ with high probability. Recall that $\E T_{\e} f_0 ~=~ 1$, and note that $Var\Big(T_{\e} f_0\Big) ~=~ \sum_{S \not = 0} \widehat{T_{\e} f_0}^2(S) ~=~ O\Big(\l^2 \cdot \E f^2\Big)$. Hence, by Chebyshev's inequality, for any $0 \le \alpha < 1$ holds \\
$Pr\Big\{T_{\e} f_0 \le \alpha\Big\} ~\le~ O\Big(\(\l^2 \cdot \E f^2\)/(1-\alpha)^2\Big)$.

\noi Therefore, taking $\alpha ~=~ 1 - \l^{1/3}$,
\[
\E_x ~\frac{\Big(T_{\e} f_1(x)\Big)^2}{T_{\e} f_0(x)} \quad \le \quad Pr\Big\{f_0 ~\le~ 1 - \l^{1/3} \Big\} ~~+~~ \Big(1 + O\(\l^{1/3}\)\Big) \cdot \E_x \Big(T_{\e} f_1(x)\Big)^2 \quad = \]
\[
O\Big(\E f^2 \Big) \cdot \l^{4/3} ~~+~~  \Big(1 + O\(\l^{1/3}\)\Big) \cdot \(\(\sum_{k=1}^n \widehat{f}^2(\{k\})\) \cdot \l ~+~ O\Big(\E f^2\Big) \cdot \l^3\) \quad =
\]
\[
\(\sum_{k=1}^n \widehat{f}^2(\{k\})\) \cdot \l ~+~ O\Big(\E f^2  \cdot \l^{4/3} \Big)
\]

\item
The third summand. Note that $\E f_1 = 0$. Hence, by Lemma~1 in \cite{OSW} (where the requirement on $f$ to be boolean does not seem to be necessary) we have, for a sufficiently small $\l$ and for some absolute constant $c > 0$ that
\[
\E_x \Big(T_{\e} f_1(x)\Big)^4 ~\le~ O\(\(\E_x \Big(T_{c\cdot\e} f_1(x)\Big)^2\)^2\) ~=~ O\(\(\E f^2\)^2 \cdot \l^2 \)
\]
We can now upper bound the third summand using the Chebyshev inequality. Taking $\alpha = 1/2$ gives $\E_x ~\frac{\Big(T_{\e} f_1(x)\Big)^4}{\Big(T_{\e} f_0(x)\Big)^3} ~\le~ O\(\(\E f^2\)^2 \cdot \l^2 \)$.

\end{enumerate}

\noi Combining these estimates leads to the claim of Lemma~\ref{lem:first level}.
\eprf

\section{Proof of Theorem~\ref{thm:KC-ln}}
\label{sec:thm}
The proof of this theorem follows very closely that of Theorem~1.11 in \cite{NELAPP}. Here we briefly describe the few required modifications. This proof is not self-contained and, in particular, borrows notation and refers to claims from the proof of Theorem~1.11 in their original numbering.

\noi As in (17) in that proof, we have that
\[
Ent\Big(T_{\e} f\Big) ~\le~  \l_2 \cdot \E_{|B| = \l_2 n,1 \in B} \bigg(Ent\Big(h~|~B\Big) - Ent\Big(h~|~\{1\}\Big) \bigg)  ~~+
\]
\beqn
\label{f}
\(1 - \l_2\) \cdot \E_{|B| = \l_2 n,1 \not \in B} ~Ent\Big(h~|~B\Big) ~~+~~\E h \cdot \phi\Bigg(Ent\(\frac{h}{\E h}~ \Big |~\{1\}\),~\e_2\Bigg) ~~ + ~~ e(n)
\eeqn
Applying this bound to the function $1-f$ gives
\[
Ent\Big(T_{\e} (1-f)\Big) ~\le~  \l_2 \cdot \E_{|B| = \l_2 n,1 \in B} \bigg(Ent\Big((1-h)~|~B\Big) - Ent\Big((1-h)~|~\{1\}\Big) \bigg)  ~~+
\]
\beqn
\label{1-f}
\(1 - \l_2\) \cdot \E_{|B| = \l_2 n,1 \not \in B} ~Ent\Big((1-h)~|~B\Big) ~~+~~\E (1-h) \cdot \phi\Bigg(Ent\(\frac{1-h}{\E(1- h)}~ \Big |~\{1\}\),~\e_2\Bigg) ~~ + ~~ e(n)
\eeqn

\noi The following three claims, which upperbound the summands on the RHS of (\ref{f}), are given, correspondingly, by Lemmas~5.1~and~5.2, and as an intermediary step in the proof of Lemma~5.3 in \cite{NELAPP}.

\begin{enumerate}
\item
\[
\E_{|B| = \l_2 n,~1 \in B} ~\bigg(Ent\Big(h~|~B\Big) - Ent\Big(h~|~\{1\}\Big) \bigg) \quad \le \quad O\Bigg(\l_2 \cdot \gamma ~+~ \gamma^2 \ln\(\frac{1}{\gamma}\)\Bigg) ~+~ e(n)
\]

\item
\[
\E_{|B| = \l_2 n,~1 \not \in B} ~Ent\Big(h~|~B\Big) \quad \le \quad O\Bigg(\l^2_2 \cdot \gamma ~+~ \l_2 \cdot \gamma^2 \ln\(\frac{1}{\gamma}\)\Bigg)
\]

\item
\[
\phi\Bigg(Ent\(\frac{h}{\E h}~ \Big |~\{1\}\),~\e_2\Bigg) \quad \le \quad \Big(1 - H(\e)\Big) ~-~ \Omega(\l \cdot \alpha) ~+~ e(n)
\]

\end{enumerate}

\noi We now deal with (\ref{1-f}). Repeating the argument, with the necessary (minor) differences, leads to the same first two bounds:
\[
\E_{|B| = \l_2 n,~1 \in B} ~\bigg(Ent\Big((1-h)~|~B\Big) - Ent\Big((1-h)~|~\{1\}\Big) \bigg) ~\le~ O\Bigg(\l_2 \cdot \gamma ~+~ \gamma^2 \ln\(\frac{1}{\gamma}\)\Bigg) ~+~ e(n)
\]
and
\[
\E_{|B| = \l_2 n,~1 \not \in B} ~Ent\Big((1-h)~|~B\Big) \quad \le \quad O\Bigg(\l^2_2 \cdot \gamma ~+~ \l_2 \cdot \gamma^2 \ln\(\frac{1}{\gamma}\)\Bigg)
\]
Indeed, this should not be surprising since, roughly speaking, these two bounds for $h$ are obtained by analysing the behavior of the squares of non-trivial Fourier coefficients of this function; and this behavior is the same for $h$ and for $1-h$.

\noi As to the third bound above, it is replaced, following the argument in the proof of Lemma~5.3, by
\[
\phi\Bigg(Ent\(\frac{1-h}{\E(1-h)}~ \Big |~\{1\}\),~\e_2\Bigg) \quad \le \quad \Big(1 - H(\e)\Big) - \Omega(\l \cdot \gamma) + e(n)
\]

\noi Combining all the bounds above gives
\[
Ent\Big(T_{\e} f\Big) ~+~ Ent\Big(T_{\e} (1-f) \Big)  ~~\le~~ \Big(1 - H(\e)\Big)  ~-~ \Omega\Big(\l \cdot \gamma \Big) ~+~ o_{\l, \gamma \rarrow 0} \Big(\l \cdot \gamma \Big) ~+~ e(n)
\]

\noi Since $\l, \gamma ~\le~ O(\delta)$, this implies that for a sufficiently small $\delta > 0$ holds
\[
Ent\Big(T_{\e} f\Big) ~+~ Ent\Big(T_{\e} (1-f) \Big)  ~~\le~~ \Big(1 - H(\e)\Big) ~+~ e(n)
\]

\noi The error term can be removed by a direct product argument (see \cite{NELAPP}), completing the proof of the theorem.

\eprf

\end{document}